\documentclass[12pt]{article}

\usepackage{amssymb,amsfonts,amsmath}
\usepackage{graphicx} 
\usepackage{indentfirst}

\parskip=\medskipamount

\newcommand {\cL}{{\cal L}}
\newcommand {\cM}{{\cal M}}
\newcommand {\cN}{{\cal N}}

\newcommand {\cR}{{\cal R}}

\def\a{\alpha}
\def \bi{\bibitem}

\def\d{\delta}

\def\G{\Gamma}

\def\m{\mu}

\def\q{\theta}

\def\s{\sigma}

\def\F{\Phi}
\def\J{\Psi}

\def\U{\Upsilon}

\newcommand{\pa}{\partial}                           

\newcommand{\be}{\begin{equation}}
\newcommand{\ee}{\end{equation}}
\newcommand{\bea}{\begin{eqnarray}}
\newcommand{\eea}{\end{eqnarray}}
\newcommand{\non}{\nonumber}
\newcommand{\1}{\underline{1}}
\newcommand{\2}{\underline{2}}

\def\dt#1{{\buildrel {\hbox{\LARGE .}} \over {#1}}}    

\newcommand{\bm}[1]{\mbox{\boldmath$#1$}}

\def\double #1{#1{\hbox{\kern-2pt $#1$}}}

\begin{document}

\begin{titlepage}

\begin{flushright}
hep-th/0602050\\
February, 2006\\
\end{flushright}
\vspace{5mm}

\begin{center}
{\Large \bf  
On superpotentials for 
nonlinear sigma-models with eight supercharges }
\end{center}

\begin{center}

{\large  
Sergei M. Kuzenko\footnote{{kuzenko@cyllene.uwa.edu.au}}
} \\
\vspace{5mm}

\footnotesize{
{\it School of Physics M013, The University of Western Australia\\
35 Stirling Highway, Crawley W.A. 6009, Australia}}  
~\\

\vspace{2mm}

\end{center}
\vspace{5mm}

\begin{abstract}
\baselineskip=14pt
\noindent
Using  projective superspace techniques,
we consider 4D $\cN=2$ and 5D  $\cN=1$ gauged
supersymmetric nonlinear sigma-models 
for which the hyper-K\"ahler target space is (an open domain of the zero 
section of) the cotangent bundle of  a real-analytic K\"ahler manifold.
As in the 4D $\cN=1$ case, one may  gauge 
those holomorphic isometries
of the base K\"ahler manifold (more precisely, their 
lifting to the cotangent bundle)
which are generated by globally defined Killing potentials.
In the U(1) case, by freezing the background vector (tropical) multiplet 
to a constant value of its gauge-invariant superfield strength, 
we demonstrate the generation of a chiral 
superpotential, upon elimination of  the auxiliary superfields 
and dualisation of the complex linear multiplets into chiral ones. 
Our analysis uncovers a $\cN=2$ superspace
origin for the results recently obtained in hep-th/0601165.
\end{abstract}

\vfill
\end{titlepage}

\newpage
\setcounter{page}{1}
\renewcommand{\thefootnote}{\arabic{footnote}}
\setcounter{footnote}{0}


\noindent
Recently, Bagger and Xiong \cite{BX} have
presented a $\cN=1$ superspace formulation for 
four- and five-dimensional supersymmetric nonlinear 
sigma-models with eight supercharges. 
In particular, they have proved the existence of a 
superpotential whenever the hyper-K\"ahler  target manifold
possesses a tri-holomorphic isometry. 
Although the corresponding potential  was previously obtained 
in components\footnote{For nonlinear sigma-models
with eight supersymmetries, 
it  follows from the famous 2D $\cN=4$ analysis of 
\cite{A-GF} that  the presence of a potential 
triggers the appearance of a central charge in the supersymmetry algebra,
with the central charge (the potential)
being proportional to (the square of)
a tri-holomorphic Killing vector field.
This is why such sigma-models are often called massive.
The structure of potentials in  2D $\cN=1,2$ 
supersymmetric sigma-models was studied in \cite{A-GF}, 
and these results were also re-cast in superfield form in  
 \cite{Gates}. }
\cite{GTT}, the superspace treatment is clearly 
quite  enlightening in several respects.

The aim of the present note is to uncover 
a manifestly supersymmetric origin\footnote{Only four supersymmetries 
are kept manifest  in the $\cN=1$ superspace formulation of \cite{BX}.}
for the results in \cite{BX}.
There are two general approaches to keep manifest 
eight supersymmetries: harmonic superspace 
(see \cite{GIOS} for a review)
and projective superspace  (see \cite{projective1, projective2}
and references therein). They are related and complementary to
each other \cite{Kuzenko}. Projective superspace used here is ideally 
suited if one is interested in re-casting the $\cN=2$ results in terms of 
$\cN=1$ superfields. Specifically, this approach  
allows one to keep $\cN=2$  supersymmetry under control 
without leaving $\cN=1$ superspace (an example of 
superspace holography).

Our starting point will be the 4D $\cN=2$ and 5D  $\cN=1$ 
supersymmetric nonlinear sigma-models studied in 
\cite{Kuzenko,GK,KL}.  
Although embracing 
only a subclass  
in  the  general family of supersymmetric actions
for self-interacting polar  multiplets  \cite{projective1}, 
these models are especially interesting in the sense that  
they naturally extend the general
4D $\cN=1$  supersymmetric nonlinear sigma-model
\cite{Zumino} 
\bea
\int  {\rm d}^4 \q \,
K \big( \F^I  , \bar{\F}^{ \bar J}   \big)~,\quad
 \qquad I,{\bar J} =1,\dots ,n~,
\label{kahlermodel}
\eea
with $K(\F ,\bar \F)$ the K\"ahler potential of a K\"ahler 
manifold $\cM$ of complex dimension $n$. 
The extension consists of the two steps: \\
(i) replace the chiral $\F$ and antichiral $\bar \F$ 
dynamical variables
with  so-called arctic $\U(w) $ and antarctic 
$\breve{\U}(w) $ projective multiplets \cite{projective1,projective2}
\be
\U (w) = \sum_{n=0}^{\infty} \U_n  \,w^n~, \qquad
\breve{\U} (w) = \sum_{n=0}^{\infty} (-1)^n {\bar \U}_n \,
\frac{1}{w^n} ~,\qquad 
w \in {\mathbb C} \setminus \{ 0 \}~;
\label{pm0}
\ee
(ii) replace the Lagrangian (\ref{kahlermodel}) with
\bea
\cL = \oint \frac{{\rm d}w}{2\pi {\rm i}w} 
\int  {\rm d}^4 \q \,
K \big( \U  , \breve{\U}   \big)\Big|~, 
\label{nact}
\eea
with the contour around the origin.
The $\U(w) $ and  $\breve{\U} (w)$ are 4D $\cN=2$ or 5D  $\cN=1$ 
superfields obeying the constraints 
\be 
\nabla_{\hat \a} (w) \U(w) =\nabla_{\hat \a} (w) \breve{\U}(w)=0~,
\label{polarconstraints}
\ee
where 
\bea
\nabla_{\hat \a} (w) 
= \left(
\begin{array}{c}
\nabla_\a (w) \\
{\bar \nabla}^{\dt \a}  (w)  
\end{array}
\right)~, 
\quad \nabla_\a (w) \equiv  w D^{\1}_\a - D^{\2}_\a ~,
\quad
{\bar \nabla}^{\dt \a} (w) \equiv {\bar D}^{\dt \a}_{ \1} + 
w {\bar D}^{\dt \a}_{ \2}~,
\label{nabla}
\eea
and $D^i_{\hat \a}= (D^i_\a , {\bar D}^{\dt \a i}) $
are the 4D $\cN=2$ or 5D  $\cN=1$
spinor covariant derivatives (see \cite{KL}
for our 5D conventions).
The constraints imply  that the dependence
of the component superfields 
$\U_n$ and ${\bar \U}_n$
on $\q^\a_{\2}$ and ${\bar \q}^{\2}_{\dt \a}$ 
is uniquely determined in terms 
of their dependence on $\q^\a_{\1}$
and ${\bar \q}^{\1}_{\dt \a}$.  In other words, 
the projective superfields depend effectively 
on half the Grassmann variables which can be chosen
to be the spinor  coordinates of 4D $\cN=1$ superspace
\be
\q^\a = \q^\a_{\1} ~, \qquad {\bar \q}_{\dt \a}=
{\bar \q}_{\dt \a}^{\1}~.
\label{theta1}
\ee 
As a result, one can deal  with reduced  superfields
$U | $, $ D^{\2}_\a U|$, $ {\bar D}_{\2}^{\dt \a} U|, \dots$
and 4D $\cN=1$ spinor covariant derivatives $D_\a$ 
and ${\bar D}^{\dt \a}$ defined 
via the bar-projection:
\be 
U| = U(x, \q^\a_i, {\bar \q}^i_{\dt \a})
\Big|_{ \q_{\2} = {\bar \q}^{\2}=0 }~,
\qquad D_\a = D^{\1}_\a \Big|_{\q_{\2} ={\bar \q}^{\2}=0} ~, 
\qquad
{\bar D}^{\dt \a} = {\bar D}_{\1}^{\dt \a} 
\Big|_{\q_{\2} ={\bar \q}^{\2}=0}~.
\label{N=1proj}
\ee 
${}$For the leading components 
$ \U_0 | =\F$ and $\U_1 | =\G$, the constraints 
(\ref{polarconstraints}) give
\bea
 {\bar D}^{\dt \a} \, \F &=&0~,
\qquad
-\frac{1}{ 4} {\bar D}^2 \, \G 
= 0~,\qquad \quad \quad \,{\rm D}=4~; \non \\ 
{\bar D}^{\dt \a} \, \F &=&0~,
\qquad
-\frac{1}{ 4} {\bar D}^2 \, \G 
= \pa_5\,  \F~,  \qquad \quad {\rm D}=5~.
\label{pm-constraints0}
\eea 
The action functional generated by the Lagrangian $\cL$, 
eq. (\ref{nact}), can be shown to follow from 
a manifestly supersymmetriic action \cite{projective2,KL}. 

The supersymmetric sigma-model (\ref{nact})
respects all the geometric features of
its 4D $\cN=1$ predecessor (\ref{kahlermodel}).
 The K\"ahler invariance of (\ref{kahlermodel})
\be
K(\F, \bar \F) \quad \longrightarrow \quad K(\F, \bar \F) ~+~ 
\Big( \Lambda(\F)
\,+\,  {\bar \Lambda} (\bar \F) \Big)
\label{kahlerinv}
\ee
turns into 
\be
K(\U, \breve{\U})  \quad \longrightarrow \quad K(\U, \breve{\U}) ~+~
\Big(\Lambda(\U) \,+\, {\bar \Lambda} (\breve{\U} ) \Big)
\ee
for the model (\ref{nact}). 
A holomorphic reparametrization $\F^I
\mapsto
f^I \big( \F \big)$ of the K\"ahler manifold has the following
counterparts
\be
\U^I (w) \quad  \mapsto  \quad f^I \big (\U(w) \big)
\ee
in the 4D $\cN=2$ and 5D $\cN=1$ cases, respectively. 
Therefore, the physical
superfields 
\be
\U^I (w)\Big|_{w=0} ~=~ \F^I ~,\qquad  \quad
 \frac{ {\rm d} \U^I (w) }{ {\rm d} w} \Big|_{w=0} ~=~ \G^I ~,
\label{geo3} 
\ee
should be regarded, respectively, as a coordinate of the K\" ahler
manifold and a tangent vector at point $\F$ of the same manifold. 
That is why the variables $(\F^I, \G^J)$ parametrize the tangent 
bundle $T\cM$ of the K\"ahler manifold $\cM$. 

The supersymmetric sigma-model (\ref{nact})
can be gauged in complete analogy with the famous gauging procedure 
for  its 4D $\cN=1$ predecessor (\ref{kahlermodel}),
the latter developed in \cite{BWitten,CL,S,HKLR,HKLR2,BWess,WB}.
We should  first recall the relevant geometric prerequisites \cite{BWitten}. 
Let $X$ and $\bar X$ be the holomorphic and antiholomorphic
parts of  a Killing vector,
\be
X= X^I(\F) \frac{\pa}{\pa \F^I} ~, 
\qquad 
\bar X = {\bar X}^{\bar I}(\bar \F) \frac{\pa}{\pa {\bar \F}^{\bar I}} ~, 
\ee
Their important properties are  \cite{BWitten}
\bea
\big[X,K\big] &=& X^I \, K_I = {\rm i} \, D ( \F , {\bar \F} ) + \eta (\F )~, \non \\
\big[ {\bar X}, K \big] &=& {\bar X}^{\bar I} \, K_{\bar I} = - {\rm i} \, D ( \F , {\bar \F} ) 
+ {\bar \eta} ({\bar \F })~,
\label{Killing}
\eea
where $D =\bar D$ is the so-called Killing potential, and $\eta$ is 
a holomorphic function. From these relations one deduces
\bea
X^I\,K_{I \bar J} =  X^I\,g_{I \bar J} =
{\rm i} \, D_{\bar J}~, \qquad
{\bar X}^{\bar I} \, K_{{\bar I}J }  =  {\bar X}^{\bar I} \, g_{{\bar I}J } =
-{\rm i} \, D_{ J}~.
\label{Killing2}
\eea

The gauging of the supersymmetric sigma-model (\ref{nact}) gives 
\bea
\cL_{\rm gauged} (V) = \oint \frac{{\rm d}w}{2\pi {\rm i}w} 
\int  {\rm d}^4 \q \,\Big\{ 
K \big( \U  , \breve{\U}  \big)
+\frac{ {\rm e}^{{\rm i}\,V L_{\bar X}} -1}{{\rm i}\,VL_{\bar X}} \, 
D(\U, \breve{\U} ) \, V
\Big\}\Big|~,
\label{nact-gauged}
\eea
with $V$ the gauge potential, and
 $L_{ \bar X} $ 
the Lie derivative along the vector ${\bar X} (\breve{\U}) $.
The gauge potential is described by the so-called tropical 
multiplet \cite{projective1,projective2,KL}
\be 
V ( w) = \sum_{n=-\infty}^{+\infty} 
V_n \,w^n~, 
\qquad \nabla_{\hat \a} (w) V(w) = 0~,
\qquad
\bar{V}_n  =  
(-1)^n \,
V_{-n}~,
\label{tropical}
\ee
possessing  the gauge freedom
\be
\d V(w) = {\rm i}\Big( \breve{\Lambda} (w)-\Lambda (w) \Big)~,
\label{lambda4}
\ee
with the gauge parameter $\Lambda (w)$ an arctic superfield.
Although our discussion is restricted to the U(1) case,
a nonabelian generalisation  is obvious. 

In 4D $\cN=2$ superspace, as a manifestation of the Higgs effect,
charged hypermultiplets can be made massive by 
freezing a U(1) vector multiplet to a constant value $\m$ of its
chiral gauge-invariant superfield strength $W$ \cite{vev,projective3}
(the same procedure also works in the 5D $\cN=1$ case, 
where the superfield strength $W$ is real).
The gauge freedom (\ref{lambda4}) can be used to bring 
a  vector multiplet of constant superfield strength 
to the form\footnote{From the point of view of supersymmetry 
without  central charge, eq. (\ref{WZgauge}) defines the Wess-Zumino gauge 
and, to preserve it, any supersymmetry transformation should be accompanied 
by  an induced gauge transformation. This generates 
a central charged supersymmetry transformation
which makes the expression for $V_0(w)$  super Poincar\'e-invariant, 
see the second reference in  \cite{vev} for more details.
See also \cite{Gates} for a similar mechanism in two dimensions.
} 
\bea
V_0 (w) &=&
-\frac{1}{w} \, \Big\{{\bar \m} \,(\q_{\1} )^2 
-  \m \,( {\bar \q}^{\2})^2 \Big\}
-2 \,  \Big\{ {\bar \m }\, \q_{\1} \q_{\2} 
+  \m \, {\bar \q}^{\1}   {\bar \q}^{\2} \Big\} \non \\
&&\,+w \,  \Big\{ 
 \m \,( {\bar \q}^{\1})^2 
 - {\bar \m } \,(\q_{\2} )^2 
\Big\}~.
\label{WZgauge}
\eea
Here the parameter $\m$ is  arbitrary  complex  if $D=4$. 
It turns out to be  real, $\m =\bar \m $, in the case $D=5$  \cite{KL}. 
It is useful to  define the arctic and antarctic components of $V_0|$:
\be
V_0 (w) | =V_{(+)} (w) +  V_{(-)} (w)~, \qquad
V_{(+)} = 
w \, \m \, {\bar \q}^2 ~, \qquad
V_{(-)}= -\frac{1}{w} \, {\bar \m}\, \q^2 ~.
\ee
Important for our consideration are the following obvious properties
\be
V_{(+)}^2 = V_{(-)}^2 = 0~.
\ee
They imply that the second term in (\ref{nact-gauged}) contains 
only two contributions
\be
\frac{ {\rm e}^{{\rm i}\,V_0 L_{\bar X}} -1}{{\rm i}\,V_0L_{\bar X}} \, D \, V_0
= V_0 \,D(\U, \breve{\U} )  
+\frac{\rm i}{2} \,V_0^2 \,[ {\bar X} (\breve{\U}), D (\U, \breve{\U} ) ]~,
\label{counterterm}
\ee
with the bar-projection assumed.

With the aid of (\ref{Killing}), the first term 
in (\ref{counterterm})
can be represented as (the bar-projection is assumed)
\bea
V_0 \,D(\U, \breve{\U} )  &=& -{\rm i}\, V_{(+)}\, \big[ X , K \big] 
+{\rm i}\, V_{(-)} \,\big[ {\bar X} , K \big] 
\non \\
&&+ {\rm i}\, V_{(+)}\, \eta (\U) 
-{\rm i}\, V_{(-)}\, {\bar \eta}(\breve{\U})~.
\eea
Here the expression in the second line does not contribute 
to the Lagrangian (\ref{nact-gauged}), 
\be
 \oint \frac{{\rm d}w}{2\pi {\rm i}w} \, V_{(+)}\, \eta (\U) 
= \oint \frac{{\rm d}w}{2\pi {\rm i}w} \, V_{(-)}\, {\bar \eta}(\breve{\U})
=0~.
\ee
With the aid of (\ref{Killing2}), the second term 
in (\ref{counterterm})
can be represented as
\be
\frac{\rm i}{2} \,V_0^2 \,[ {\bar X} (\breve{\U}), D (\U, \breve{\U} ) ] \Big|
= V_{(+)}\, V_{(-)}\, [X, [{\bar X}, K]]~.
\ee

The relations obtained allow  us to rewrite 
the Lagranian, $\cL_{\rm gauged} (V_0) $,
as follows:
\be
\cL'=
\cL_{\rm gauged} (V_0) = \oint \frac{{\rm d}w}{2\pi {\rm i}w} 
\int  {\rm d}^4 \q \,
K \big({\bm  \U}  , \breve{\bm \U}  \big)\Big|~, 
\qquad  {\bm \U}^I =  \U^I 
-{\rm i}\, V_{(+)}\,  X^I (\U)~.
\label{nact-gauged2} 
\ee
Among the component superfields of the modified arctic multiplet
\be
{\bm \U} (w) \Big|= \sum_{n=0}^{\infty} {\bm \U}_n  \,w^n~, 
\label{pm}
\ee
the leading component does not change at all, 
${\bm  \U}_0  =\F$, while  
the next-to-leading  one, $ {\bm \U}_1  ={\bm \G}$,
obeys the modified linear constraint 
\bea
 -\frac{1}{ 4} {\bar D}^2 \, {\bm \G}^I &=& -{\rm i}\,\m \, X^I(\F)~, 
\qquad \qquad \quad ~D=4~; \non   \\
-\frac{1}{ 4} {\bar D}^2 \,{\bm  \G}^I 
&=& -{\rm i}\,\m \, X^I(\F)
+ \pa_5\,  \F^I~, \qquad D=5~. 
\label{pm-constraints}
\eea 
The expressions appearing in both sides of these relations
transform as holomorphic target-space vectors.
It is worth pointing out that such generalised constraints 
for 4D $\cN=1$ complex linear 
superfields were designed many years ago \cite{DG}.

The auxiliary superfields ${\bm \U}_2, {\bm \U}_3, \dots$, and their
conjugates,  can be eliminated  with the aid of the 
corresponding algebraic equations of motion
\be
 \oint {{\rm d} w} \,w^{n-1} \, 
 \frac{\pa K({\bm \U}, \breve{\bm \U} ) }{\pa {\bm \U}^I} = 0~,
\qquad n \geq 2 ~.               
\label{int}
\ee
Their elimination  can be  carried out
using the ansatz\footnote{It is explained in \cite{GK} 
how to eliminate the auxiliary superfields in the case 
of symmetric K\"ahler spaces, and the example of 
$\cM ={\mathbb C}P^n$ is explicitly elaborated.} 
\cite{Kuzenko2}
\bea
{\bm \U}^I_n = \sum_{p=o}^{\infty} 
G^I{}_{J_1 \dots J_{n+p} \, \bar{L}_1 \dots  \bar{L}_p} (\F, {\bar \F})\,
{\bm \G}^{J_1} \dots {\bm \G}^{J_{n+p}} \,
{\bar {\bm \G}}^{ {\bar L}_1 } \dots {\bar {\bm \G}}^{ {\bar L}_p }~, 
\qquad n\geq 2~.
\eea
Upon elimination of the auxiliary superfields,
the Lagrangian (\ref{nact-gauged2}) 
takes the form
\bea
\cL_{\rm tb}(\F, \bar \F, {\bm \G}, \bar {\bf \G})  
&=& 
\int 
{\rm d}^4 \q \, \Big\{\,
K \big( \F, \bar{\F} \big) - g_{I \bar{J}} \big( \F, \bar{\F} 
\big) {\bm \G}^I {\bar {\bm \G}}^{\bar{J}} 
\non\\
&&\qquad +
\sum_{p=2}^{\infty} \cR_{I_1 \cdots I_p {\bar J}_1 \cdots {\bar 
J}_p }  \big( \F, \bar{\F} \big) {\bm \G}^{I_1} 
\dots {\bm \G}^{I_p} {\bar {\bm \G}}^{ {\bar J}_1 } 
\dots {\bar {\bm \G}}^{ {\bar J}_p }~\Big\}~, 
\eea
where the tensors $\cR_{I_1 \cdots I_p {\bar J}_1 \cdots {\bar 
J}_p }$ are functions of the Riemann curvature $R_{I {\bar 
J} K {\bar L}} \big( \F, \bar{\F} \big) $ and its covariant 
derivatives.  Each term in the action contains equal powers
of $\bm \G$ and $\bar {\bm \G}$, since the original model 
 (\ref{nact-gauged2}) 
is invariant under rigid U(1)  transformations
\be
{\bm \U}(w) ~~ \mapsto ~~ {\bm \U}({\rm e}^{{\rm i} \a} w) 
\quad \Longleftrightarrow \quad 
{\bm \U}_n(z) ~~ \mapsto ~~ {\rm e}^{{\rm i} n \a} {\bm \U}_n(z) ~.
\label{rfiber}
\ee

${}$For the theory with Lagrangian  
$\cL_{{\rm tb}}(\F, \bar \F, {\bm \G}, \bar {\bm \G})  $, 
we can develop a dual formulation involving only chiral superfields 
and their  conjugates as the dynamical variables. 
In the five-dimensional case, for concreteness,
consider the first-order action 
\bea
&&
S_{\rm tb} 
- \int {\rm d}^5 x \,\Big\{ 
\int {\rm d}^2 \q\, \J_I \Big( \pa_5\,  \F^I 
-{\rm i}\,\m \, X^I(\F)
+
\frac{1}{ 4} {\bar D}^2 \, {\bm \G}^I \Big)
~+~{\rm c.c.} \Big\} \non \\
&=&
S_{\rm tb} 
+ \int {\rm d}^5 x \,\Big\{ 
\int {\rm d}^4 \q \, \J_I \,{\bm \G}^I
+\int {\rm d}^2 \q\, \J_I \Big( 
{\rm i}\,\m \, X^I(\F)
-\pa_5\,  \F^I \Big)
~+~{\rm c.c.} \Big\} ~,
\non
\eea
where 
$S_{\rm tb} =\int {\rm d}^5 x\,\cL_{{\rm tb}}(\F, \bar \F, {\bm \G}, \bar {\bm \G})$,
and the tangent vector ${\bm \G}^I$ is now  complex unconstrained, 
while the one-form $\J_I$ is chiral, ${\bar D}_{\dt \a} \J_I =0$.
Upon elimination of $\bm \G$ and $\bar {\bm \G}$, 
with the aid of their equations of 
motion, the action turns into $S_{{\rm cb}}[\F, \bar \F, \J, \bar \J]$.
Its target space is  the cotangent 
bundle $T^*\cM$ of the K\"ahler manifold $\cM$. 

Let us consider the superpotential term in four dimensions 
\be
{\rm i}\,\m 
\int {\rm d}^2 \q\, \J_I 
\, X^I(\F) 
= {\rm e}^{{\rm i}\s} \int {\rm d}^2 \q\, \J_I  \,\tilde{X}^I(\F) ~,
\qquad  \tilde{X}^I(\F) = |\m|\,{X}^I(\F) ~.
\ee
In terms of the redefined holomorphic Killing vector $ \tilde{X}^I(\F) $,
the superpotential is defined uniquely up to a phase factor, 
${\rm e}^{{\rm i} \s} = {\rm i}\,\m / |\m|$, and this
agrees with \cite{BX}.

Massive supersymmetric sigma-models on the cotangent bundles
of complex projective spaces  ${\mathbb C}P^n$ and
Grassmannians $G(k,n)$ possess interesting topological solutions,
see \cite{GTT2,ANNS} and references therein. 
It would be interesting to extend this study to 
more general target spaces $T^*\cM$.
Our work provides a constructive approach 
to generate such sigma-models.

\noindent 
{\it Note added}: The author has been informed that a
similar construction for susy nonlinear sigma-models in 6D is
presently under investigation \cite{GPT}.
\vskip.5cm

\noindent
{\bf Acknowledgements:}\\
It is a pleasure to thank Ian McArthur and Shane McCarthy for reading the manuscript, and Jim Gates 
for comments. This work  is supported  
by the Australian Research Council and by a UWA research grant.

\small{

}

\end{document}